\documentclass[reprint,preprintnumbers,amsmath,amssymb,aps,prl,superscriptaddress]{revtex4-2}
\usepackage{graphicx}
\usepackage{dcolumn}
\usepackage{bm}
\usepackage{hyperref}
\usepackage{siunitx}
\usepackage[mathlines]{lineno}
\usepackage{float}

\newcommand{\cv}{\text{CV}}

\newcommand{\Cov}{\text{Cov}}
\newcommand{\bphi}{\boldsymbol{\phi}}

\begin{document}
\preprint{FERMILAB-PUB-26-0808-PPD}
\title{Measurements of the total and differential inclusive electron neutrino charged-current cross sections on argon with SBND}

\newcommand{\ABC}{Universidade Federal do ABC, Santo Andre, SP 09210-580, Brazil}
\newcommand{\ANL}{Argonne National Laboratory, Lemont, IL 60439, USA}
\newcommand{\Alfenas}{Universidade Federal de Alfenas, Pocos de Caldas, MG 37715-400, Brazil}
\newcommand{\BNL}{Brookhaven National Laboratory, Upton, NY 11973, USA}
\newcommand{\Bern}{Universit{\"a}t Bern, Bern CH-3012, Switzerland}
\newcommand{\CIEMAT}{Centro de Investigaciones Energ{\'e}ticas, Medioambientales y Tecnol{\'o}gicas, Madrid E-28040, Spain}
\newcommand{\CSU}{Colorado State University, Fort Collins, CO 80523, USA}
\newcommand{\CTI}{Centro de Tecnologia da Informa{\c{c}}{\~a}o Renato Archer, Campinas, SP 13069-901, Brazil}
\newcommand{\Campinas}{Universidade Estadual de Campinas, Campinas, SP 13083-970, Brazil}
\newcommand{\Chicago}{University of Chicago, Chicago, IL 60637, USA}
\newcommand{\Columbia}{Columbia University, New York, NY 10027, USA}
\newcommand{\Edinburgh}{University of Edinburgh, Edinburgh EH9 3FD, United Kingdom}
\newcommand{\FNAL}{Fermi National Accelerator Laboratory, Batavia, IL 60510, USA}
\newcommand{\Florida}{University of Florida, Gainesville, FL 32611, USA}
\newcommand{\GeorgiaTech}{Georgia Institute of Technology, Atlanta, GA 30332, USA}
\newcommand{\Granada}{Universidad de Granada, Granada E-18071, Spain}
\newcommand{\Hanyang}{Hanyang University, 222, Wangsimni-ro, Seongdong-gu, Seoul, 04763, Korea}
\newcommand{\Holyoke}{Mount Holyoke College, South Hadley, MA 01075, USA}
\newcommand{\IIS}{Indian Institute of Science, Centre for High Energy Physics, Bangalore 560012, India}
\newcommand{\IIT}{Illinois Institute of Technology, Chicago, IL 60616, USA}
\newcommand{\Imperial}{Imperial College London, London SW7 2AZ, United Kingdom}
\newcommand{\Kansas}{University of Kansas, Lawrence, KS 66045, USA}
\newcommand{\LANL}{Los Alamos National Laboratory, Los Alamos, NM 87545, USA}
\newcommand{\LSU}{Louisiana State University, Baton Rouge, LA 70803, USA}
\newcommand{\Lancaster}{Lancaster University, Lancaster LA1 4YW, United Kingdom}
\newcommand{\Liverpool}{University of Liverpool, Liverpool L69 7ZE, United Kingdom}
\newcommand{\Manchester}{University of Manchester, Manchester M13 9PL, United Kingdom}
\newcommand{\Michigan}{University of Michigan, Ann Arbor, MI 48109, USA}
\newcommand{\Minnesota}{University of Minnesota, Minneapolis, MN 55455, USA}
\newcommand{\NotreDame}{University of Notre Dame, Notre Dame, IN 46556, USA}
\newcommand{\Oxford}{University of Oxford, Oxford OX1 3RH, United Kingdom}
\newcommand{\Palermo}{Universit{\`a} degli Studi di Palermo, Palermo I-90123, Italy}
\newcommand{\Penn}{University of Pennsylvania, Philadelphia, PA 19104, USA}
\newcommand{\QueenMary}{Queen Mary University of London, London E1 4NS, United Kingdom}
\newcommand{\Rutgers}{Rutgers University, Piscataway, NJ 08854, USA}
\newcommand{\SaoJose}{Instituto Tecnol{\'o}gico de Aeron{\'a}utica S{\~a}o Jos{\'e} dos Campos, SP 12228-900, Brazil}
\newcommand{\Sheffield}{University of Sheffield, School of Mathematical and Physical Sciences, Sheffield S3 7RH, United Kingdom}
\newcommand{\Sussex}{University of Sussex, Brighton BN1 9RH, United Kingdom}
\newcommand{\Syracuse}{Syracuse University, Syracuse, NY 13244, USA}
\newcommand{\TAMU}{Texas A{\&}M University, College Station, TX 77843, USA}
\newcommand{\Tufts}{Tufts University, Medford, MA 02155, USA}
\newcommand{\UCSB}{University of California Santa Barbara, Santa Barbara, CA 93106, USA}
\newcommand{\UTA}{University of Texas at Arlington, Arlington, TX 76019, USA}
\newcommand{\VirginiaTech}{Virginia Tech, Center for Neutrino Physics, Blacksburg, VA 24060, USA}
\newcommand{\Warwick}{University of Warwick, Coventry CV4 7AL, United Kingdom}
\affiliation{\ABC}
\affiliation{\ANL}
\affiliation{\Alfenas}
\affiliation{\BNL}
\affiliation{\Bern}
\affiliation{\CIEMAT}
\affiliation{\CSU}
\affiliation{\CTI}
\affiliation{\Campinas}
\affiliation{\Chicago}
\affiliation{\Columbia}
\affiliation{\Edinburgh}
\affiliation{\FNAL}
\affiliation{\Florida}
\affiliation{\GeorgiaTech}
\affiliation{\Granada}
\affiliation{\Hanyang}
\affiliation{\Holyoke}
\affiliation{\IIS}
\affiliation{\IIT}
\affiliation{\Imperial}
\affiliation{\Kansas}
\affiliation{\LANL}
\affiliation{\LSU}
\affiliation{\Lancaster}
\affiliation{\Liverpool}
\affiliation{\Manchester}
\affiliation{\Michigan}
\affiliation{\Minnesota}
\affiliation{\NotreDame}
\affiliation{\Oxford}
\affiliation{\Palermo}
\affiliation{\Penn}
\affiliation{\QueenMary}
\affiliation{\Rutgers}
\affiliation{\SaoJose}
\affiliation{\Sheffield}
\affiliation{\Sussex}
\affiliation{\Syracuse}
\affiliation{\TAMU}
\affiliation{\Tufts}
\affiliation{\UCSB}
\affiliation{\UTA}
\affiliation{\VirginiaTech}
\affiliation{\Warwick}
\author{L.~Aliaga-Soplin} \affiliation{\UTA}
\author{D.~Andrade Aldana} \affiliation{\IIT}
\author{C.~Andreopoulos} \affiliation{\Liverpool}
\author{M.~Andriamirado} \affiliation{\Kansas}
\author{A.~Antonakis} \affiliation{\UCSB}
\author{S.~Balasubramanian} \affiliation{\Holyoke}
\author{A.~Barnard} \affiliation{\Oxford}
\author{V.~Basque} \affiliation{\Edinburgh}
\author{J.\,R.~Bateman} \affiliation{\Manchester}\affiliation{\Imperial}
\author{A.~Beever} \affiliation{\Sheffield}
% ...
\collaboration{The SBND Collaboration}
\thanks{sbnd\_info@fnal.gov}
\noaffiliation

\begin{abstract}
This work presents measurements of the total and single-differential electron neutrino charged-current cross sections on argon using the Short-Baseline Near Detector (SBND) at Fermilab. We present the highest statistics inclusive $\nu_e + \bar \nu_e$ (95.2\% $\nu_e$ and 4.8\% $\bar\nu_e$) charged-current cross section measurement on argon to date and the first inclusive $\nu_e + \bar \nu_e$ cross section measurement using the Booster Neutrino Beam. Direct measurements of this channel are crucial for future studies of both long- and short-baseline neutrino oscillations through electron neutrino appearance, which in turn are vital for answering open questions in neutrino phenomenology and searching for beyond the Standard Model physics at the intensity frontier. The data used for this work were collected with an exposure of $8.81 \times10^{19}$ protons on target, and 729 candidate events are selected with a minimum electron energy of 500\,MeV. The single-differential cross sections are extracted in electron energy and angle with respect to the beam axis. The flux-averaged total cross section is measured to be $\sigma = [3.90 \pm 0.18\,\text{(stat)} \pm 0.41\,\text{(syst)}] \times 10^{-39}$ \unit{\square{\centi\meter}\per nucleon} for a predicted average neutrino energy of $\langle E_\nu \rangle=1.8$\,GeV for selected events. The total and differential cross sections are compared to predictions from common neutrino event generators. The data show a preference for generators that predict a lower overall cross section. 
\end{abstract}
\maketitle
\paragraph{Introduction}
Electron neutrino ($\nu_e$) appearance in a muon neutrino ($\nu_\mu$) source is an essential neutrino oscillation channel for answering remaining questions in the field of neutrino physics~\cite{Huber:2022lpm}. For long-baseline experiments, $\nu_e$ appearance measurements enable the determination of the neutrino mass ordering and the search for charge-parity ($CP$) violation in the neutrino sector~\cite{DUNE:2020jqi, Hyper-Kamiokande:2018ofw, NOvA:2025tmb, T2K:2023smv, T2K:2025wet}. For short-baseline experiments, this channel provides sensitivity to a broad class of physics beyond the Standard Model~\cite{Arguelles:2019xgp}, including light singlet fermions~\cite{Acero:2022wqg, Abdullahi:2022jlv} and hidden-sector portals motivated by the absence of a Standard Model mechanism for neutrino mass~\cite{Batell:2016zod}. 
To enable current- and next-generation experiments to robustly explore these open questions in neutrino physics, it is critical to model $\nu_e$-nucleus interactions with high precision.
Differences between $\nu_\mu$ and $\nu_e$ cross sections due to the lepton mass dependence and radiative corrections~\cite{Day:2012gb, Nikolakopoulos:2019qcr} may lead to additional uncertainties on the $\nu_e$/$\nu_\mu$ ratio, reducing the sensitivity for neutrino oscillation and $CP$ violation searches and motivating direct measurements of $\nu_e$ interactions. Although the number of direct $\nu_e$ measurements has increased in recent years~~\cite{T2K:2014lbi,MINERvA:2015jih,T2K:2020lrr,NOvA:2022see,T2K:2025smz}, particularly on argon~\cite{ArgoNeuT:2020kir,MicroBooNE:2021gfj,MicroBooNE:2021ppm,MicroBooNE:2022tdd,MicroBooNE:2025pvb,MicroBooNE:2025aiw,MicroBooNE:2026cnl}, many of these results have limited statistical power, restricting the understanding of the agreement between $\nu_e$ cross section models and data. 

The Short-Baseline Near Detector (SBND) is the near detector for the Short-Baseline Neutrino Program at the Fermi National Accelerator Laboratory (Fermilab)~\cite{Machado:2019oxb,SBND:2026tee}. Only 110\,meters away from the Booster Neutrino Beam (BNB) target, SBND observes a high neutrino flux, making it well-poised to explore complex particle and nuclear physics for neutrino interactions at the few-GeV scale with high statistical precision. These measurements will be critical for neutrino interaction modeling and future neutrino oscillation measurements~\cite{DUNE:2020jqi}. 

This work presents measurements of the total and differential inclusive $\nu_e+\bar\nu_e$ charged-current (CC) cross sections for events with an outgoing lepton with energy greater than \qty{500}{\mega\eV}. The 500\,MeV threshold is applied to reduce potential sensitivity to the region where the MiniBooNE experiment observed an excess of low-energy, electron-like events~\cite{MiniBooNE:2020pnu}. For these measurements, we assume no (or negligible) short-baseline oscillations~\cite{ParticleDataGroup:2026mpi}. As electrons and positrons are indistinguishable in SBND, they are collectively referred to as electrons for the remainder of this work. This work presents the highest statistics measurement of charged-current $\nu_e+\bar\nu_e$ interactions on argon to date, and is the first inclusive $\nu_e+\bar\nu_e$ cross section measurement performed on the BNB at Fermilab. Due to the focusing configuration of the BNB and the lower anti-neutrino interaction rate, simulations indicate the selected signal interactions are 95.2\% $\nu_e$. Furthermore, this measurement has substantial sensitivity to the $>$1\,GeV region, with over 85\% (30\%) of the signal interactions predicted to have neutrino energy above 1 (2)\,GeV. Thus, this $\nu_e$-Ar measurement has significant overlap with the first oscillation maximum at DUNE ~\cite{DUNE:2020ypp}.

\paragraph{The SBND Experiment}
The central detector components of SBND are the time projection chamber and the photon detection system, both of which are contained within a membrane cryostat and completely immersed in liquid argon~\cite{SBND:2026tee}. 
The \qty{112}{tonne} active mass liquid argon time projection chamber (LArTPC) consists of two independent drift volumes, each read out by a wire-based plane assembly that measures the current induced or collected by drifting ionization electrons. Each volume is \qty{5}{\meter} long, \qty{4}{\meter} high, and \qty{2}{\meter} in the drift direction. The photon detection system consists of 120 photomultiplier tubes (PMTs) and 192 X-ARAPUCAs mounted behind the wire planes and passive wavelength-shifting components mounted at the cathode~\cite{SBND:2024vgn,SBND:2026yvd}. During the data collection period for this work, scintillation light was collected by the PMTs only. To trigger on beam-coincident neutrino interactions, we rely on the coincidence of scintillation light detected by the PMTs and the gate corresponding to the time when neutrinos are expected to arrive at SBND. More details about the design and performance of the detector subsystems can be found in Ref.~\cite{SBND:2026tee}. 

SBND is exposed to the BNB, which is supplied with 8\,GeV protons from the Booster synchrotron at Fermilab. These protons are directed onto a beryllium target, generating a secondary beam of hadrons. These hadrons are then focused by a horn and decay to produce neutrinos. 
In neutrino mode, the BNB flux predominantly consists of $\nu_\mu$ at $\sim$92.59\%, followed by $\bar\nu_\mu$ at $\sim$5.90\%, and an intrinsic contamination from $\nu_e/\bar\nu_e$ at $\sim$0.55\% and $\sim$0.06\% respectively. The $\nu_\mu$ flux comes primarily from pion decays, whereas the $\nu_e$ flux is dominated by three-body decays of kaons and the decay of muons from pion decay~\cite{MiniBooNE:2008hfu}. The average neutrino energy from the $\nu_e/\bar\nu_e$ flux is 0.90\,GeV. The $\nu_e/\bar\nu_e$ flux is included in the data release, see Sec.~\ref{sec:sm-flux}~\cite{SBND:supplemental}. 

\begin{figure}
    \centering
    \includegraphics[width=1.0\linewidth]{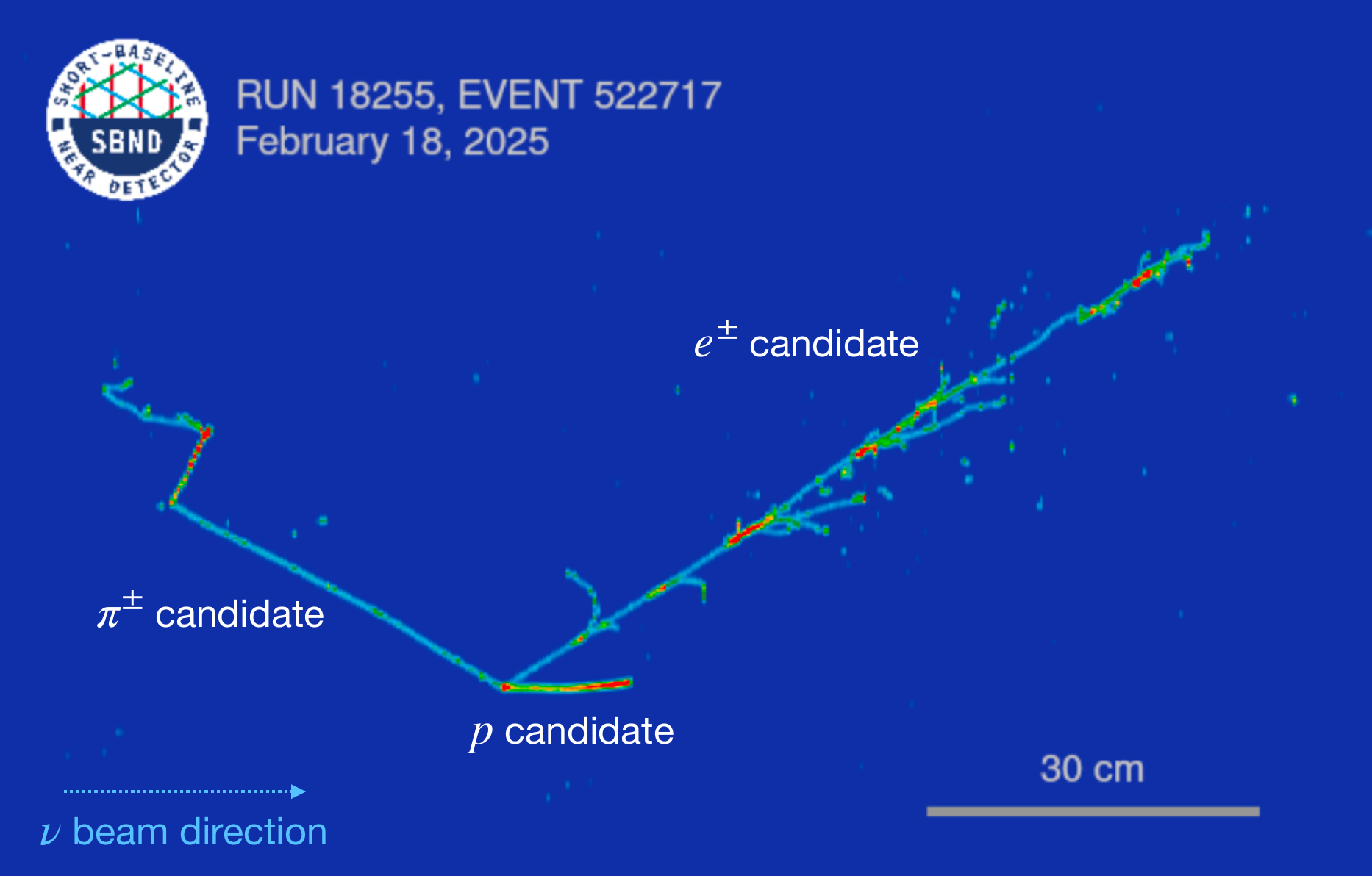}
    \caption{A selected charged-current electron (anti-)neutrino candidate. This event is characterized by an electron-like electromagnetic shower, a proton-like track, and a pion-like track exiting from a common interaction vertex.
    }
    \label{fig:evd}
\end{figure}

In this work, an exposure of \qty{8.81e19}{}\,protons on target (POT), collected between late February and early April 2025, is used. This exposure, using neutrino mode data from the BNB, is obtained after applying beam quality cuts. A $\nu_e$($\bar\nu_e$) CC interaction candidate from this dataset is shown in Fig.~\ref{fig:evd}. In addition to the on-beam data collected while BNB is running, off-beam data is also recorded to characterize the cosmic ray rate. 

\paragraph{Event Simulation and Reconstruction}
The simulation and reconstruction of neutrino interactions, which will also be referred to as \textit{events}, are performed using the \texttt{LArSoft} framework~\cite{Snider:2017wjd}. The BNB flux simulation utilizes methods originally developed for the MiniBooNE experiment~\cite{MiniBooNE:2008hfu}, but modified for the location of the SBND detector. Neutrino interactions are simulated by the \texttt{GENIE v3.4.0 Ar23\_20i\_00\_000} neutrino generator~\cite{Andreopoulos:2009rq,GENIE:2021npt} and cosmic rays are simulated with \texttt{CORSIKA}~\cite{Heck:1998vt,MicroBooNE:2020fmc, MicroBooNE:2024prh}.
\texttt{Geant4}~\cite{GEANT4:2002zbu}, the semi-analytical photon model~\cite{Garcia-Gamez:2020xrv}, and the \texttt{Wire-Cell }framework~\cite{MicroBooNE:2018swd} simulate particle propagation, photon transport, and the TPC response, respectively. The measurement presented in this work uses the \texttt{Pandora} reconstruction paradigm~\cite{Marshall:2015rfa,MicroBooNE:2017xvs} to reconstruct TPC interactions. 

To reconstruct the energy of electron- and photon-induced electromagnetic showers, we convert the charge collected on the outermost (collection) wire plane to energy via uniform scaling factors to account for electron-ion recombination and reconstruction incompleteness. The average recombination scaling factor, $R=0.64$, is estimated from simulation assuming the ellipsoidal modified box recombination model~\cite{ArgoNeuT:2013kpa, ICARUS:2024woh}. A shower incompleteness correction factor $C=1.17$ is obtained from $\pi^0$ mass peak studies on both data and simulation, similar to the methodology from Ref.~\cite{MicroBooNE:2019rgx}. Additional details on the shower energy calibration can be found in the Supplemental Material, Sec.~\ref{sec:sm-shower-cal}~\cite{SBND:supplemental}.

Spatial and temporal calibrations are applied to both the TPC and PMT data to ensure uniformity across detector subsystems.

\paragraph{Event selection}
The signal definition for this analysis requires a $\nu_e$($\bar\nu_e$) CC interaction with an outgoing electron with energy greater than \qty{500}{\mega\eV} and an interaction vertex within the fiducial volume of the detector. The fiducial volume is described in Sec.~\ref{sec:sm-flux} of the Supplemental Material~\cite{SBND:supplemental}.

To reject cosmic-ray activity and low-energy beam-induced background, we require a minimum amount of measured light from the PMTs to enforce an effective energy cut. The timing of the measured light must fall within the \qty{1.6}{\micro\second} BNB beam spill. The compatibility of the spatial distributions as well as the amount of total charge and in-time light are also evaluated to reject clear mismatches between the optical signal and TPC neutrino candidates. Following the signal definition, all remaining neutrino candidates are required to have at least one reconstructed shower with energy greater than 500 MeV and a \texttt{Pandora} reconstructed neutrino vertex within the fiducial volume. \texttt{Pandora} generates shower-like and track-like hypotheses for each particle through a boosted decision tree. To reject muon neutrino interactions, we apply a cut on the leading track length in the interaction, requiring that the length of the longest reconstructed track is less than \qty{200}{\centi\meter}. This requirement rejects less than 1\% of selected signal events with charged pions. 

After the muon rejection selection cut, the remaining dominant backgrounds are $\nu_\mu$CC and neutral current (NC) $\pi^0$ interactions. To distinguish electron-induced and photon-induced electromagnetic (EM) showers, we take advantage of the high spatial and calorimetric resolution of LArTPCs. The initial energy loss, $dE/dx$, for photon-induced EM showers corresponds to the conversion to an $e^+e^-$ pair, and thus has twice the value of that of an electron. Additionally, photons have a mean free path of tens of centimeters in liquid argon~\cite{MicroBooNE:2019rgx}, resulting in a finite conversion gap of $\mathcal{O}(\text{cm})$ before interacting with the medium. 
The reconstructed $dE/dx$ at the beginning of shower development, the conversion gap, and the transverse and longitudinal profiles of the shower are used to remove these backgrounds. A more detailed description of the selection is provided in the Supplemental Material, Secs.~\ref{sec:sm-selection} and~\ref{sec:sm-dedx}~\cite{SBND:supplemental}. 

\begin{figure*}[!htpb]
    \centering
    \includegraphics[width=\textwidth]{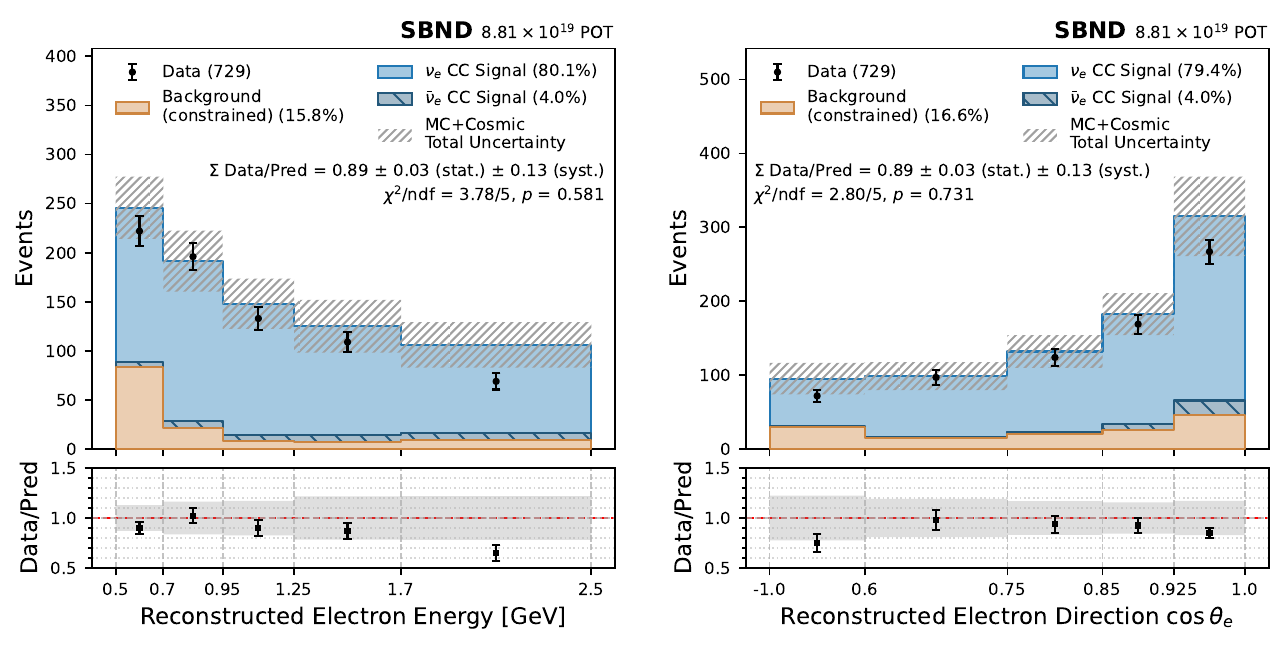}
    \caption{Data and predicted event rate in the signal region after applying the conditional covariance background constraint (CCBC) on the predicted background rate for the reconstructed electron energy (left) and the reconstructed electron direction (right). The black data points include the data statistical error, and the shaded band includes the total systematic and statistical uncertainties on the prediction.}
    \label{fig:constr_evt_rate}
\end{figure*}

We select 729 events after all selection cuts. The selection has an average efficiency of 36.7\% and a purity of 80.1\% for signal events. The selected $\nu_e+\bar\nu_e$CC events are predicted to be 55.7\% quasi-elastic, 25.5\% resonant, 12.9\% meson-exchange current, and 5.6\% deep-inelastic scattering. Of the selected signal events, 95.2\% (4.8\%) are predicted to be $\nu_e$ ($\bar\nu_e$).

\paragraph{Systematic Uncertainty Estimation} Systematic uncertainties can impact the predicted selection efficiency, bin-to-bin migration, and total event rate. Systematic uncertainties are evaluated by varying model parameters to produce alternate simulations (universes), either via event reweighting or full re-simulation.
To quantify the systematic uncertainties on our measurement, we use the covariance matrix formalism. This formalism is described in the Supplemental Material, Sec.~\ref{sec:sm-covariance}~\cite{SBND:supplemental}. The total covariance matrix is the sum of all systematic and statistical covariance matrices, where we assume Poisson uncertainties for the data statistical uncertainty.

Systematic uncertainties on the underlying cross section model for each neutrino-nucleus interaction channel as well as final state interactions are included with \texttt{GENIEReWeight} and \texttt{nusystematics}~\cite{Andreopoulos:2015wxa,NuSystematicsSoftware}. The dominant cross section uncertainties are due to the modeling of meson-exchange current interactions. For these systematics, only the impact on the selection efficiency, bin-to-bin migration, and total background event rate are included in the covariance matrix. 
To characterize the detector model uncertainty, variations in PMT response~\cite{SBND:2026tee}, recombination parameters~\cite{ICARUS:2024woh}, electric field non-uniformities, TPC wire response~\cite{MicroBooNE:2021roa}, and calorimetry corrections are implemented. Particle re-interaction uncertainties are implemented using the \texttt{Geant4Reweight} framework~\cite{Calcutt:2021zck}. The systematics on the beam flux are characterized by varying the production cross sections of hadrons and the distribution and magnitude of the current within the horn~\cite{MiniBooNE:2008hfu,MiniBooNE:2008yuf}. To quantify the uncertainty on the cosmic-ray rate, we treat the event rate of selected off-beam data as the central value prediction and compare with selected simulated cosmic rays from \texttt{CORSIKA}~\cite{Heck:1998vt} as an alternate universe, leading to a 0.5\% fractional uncertainty. For the number of target argon atoms and exposure accounting, we apply a uniform normalization uncertainty leading to a 1.2\% and 2.5\% fractional uncertainty on the cross section.

The contributions of each systematic uncertainty for the total cross section are summarized in Table~\ref{tab:systematics}. For the differential cross section measurement, the largest two contributions to the uncertainties per bin are from the flux model, dominated by modeling of $K^0$ and $K^+$ hadrons (8-14\%), and the detector model, dominated by calorimetric uncertainties on shower energy estimation (4-23\%). The bin-by-bin uncertainty breakdowns for the differential measurements are provided in the Supplemental Material, Sec.~\ref{sec:sm-syst}~\cite{SBND:supplemental}. 

\begin{table}[h]
\caption{\label{tab:systematics} Summary of systematic uncertainty contributions to the total cross section measurement. The constraint is applied to background uncertainties only.}
\begin{ruledtabular}
\begin{tabular}{lc}
\textrm{Source of uncertainty} & \multicolumn{1}{c}{\textrm{Fractional Uncertainty (\%)}} \\
\colrule
Beam flux                & 8.4 \\
Detector model           & 4.6 \\
Cross section model      & 3.8 \\ 
Exposure accounting      & 2.5 \\ 
Particle re-interactions & 1.4 \\
Number of targets        & 1.2 \\
MC + Off-beam statistics & 0.8 \\
Cosmic-ray simulation    & 0.5 \\
Data statistics          & 3.9 \\
\hline
Total (pre-constraint)   & 11.5\\ 
Total (constrained)      & 11.2\\ 
\end{tabular}
\end{ruledtabular}
\end{table}

\paragraph{Background Constraint}
As this measurement targets electromagnetic showers greater than \qty{500}{\mega\eV},
the background is primarily photons originating from the decay of high momentum neutral pions ($\pi^0$), with $\langle p_{\pi^0} \rangle=$ \qty{0.74}{\giga eV \per c}. In contrast to the $\nu_e$ signal, the background is dominated by deep-inelastic scattering (42.8\%) and resonant (42.1\%) interactions. To perform a data-driven constraint, we designed a sideband to target the high-momentum $\pi^0$ background. The conditional covariance background constraint (CCBC) procedure ~\cite{Gardiner:2024gdy} is used to constrain the shape, normalization, and systematic uncertainties of the predicted background event rate.

The sideband selection follows similar cosmic rejection criteria as the signal selection, but selects photon candidates based on the leading shower $dE/dx$ and conversion gap. To include contributions from $\nu_\mu$ interactions, no track length cut is applied. The event rate distributions for selected sideband events are provided in the Supplemental Material, Sec.~\ref{sec:sm-sideband}~\cite{SBND:supplemental}. 

The CCBC updates the background prediction while leaving the signal prediction untouched. As such, the constraint reduces the uncertainty on the total background event rate from 20.4\% to 10.5\%, whereas the uncertainty on the total background-subtracted event rate is reduced from 11.5\% to 11.2\%. The constraint reduces the total number of predicted background events from 166 to 132, corresponding to a background reduction of 20.5\%. The predicted selection purity is thus updated to 84.5\%. This large normalization shift suggests potential mis-modeling of the production of high momentum neutral pions in our cross section model.

The constrained, predicted event rate and the observed data are shown in Fig.~\ref{fig:constr_evt_rate}. For the electron energy, the highest energy bin also serves as an overflow bin. The unconstrained event rate is provided in the Supplemental Material, Sec.~\ref{sec:sm-unconstr}~\cite{SBND:supplemental}. After the constraint is implemented, uncertainties are no longer subdivided by category and thus a single total uncertainty per bin, along with bin-to-bin correlations, is obtained. 

\paragraph{Unfolding}
\begin{figure*}[!htpb]
    \centering
    \includegraphics[width=\textwidth]{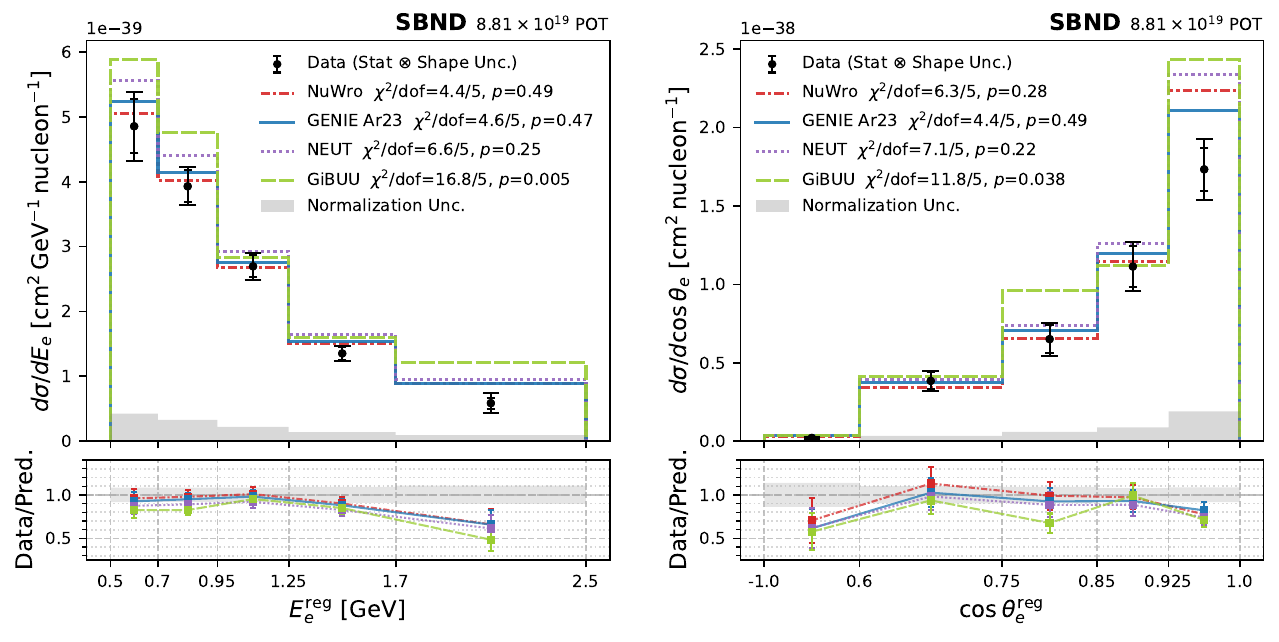}
    \caption{Flux-averaged differential cross sections as a function of outgoing electron energy (left) and outgoing electron angle (right). The black data points show the unfolded result with statistical (inner error bar) and shape-only systematic (outer error bar) uncertainties combined. The bottom data-prediction ratio error bars include the combined statistical and shape-only uncertainties. The shaded gray region shows the per-bin normalization uncertainty for top and bottom plots. Generator predictions are smeared with the regularization matrix $A_C$.}
    \label{fig:unfolded_xsec}
\end{figure*}

The total cross section $\sigma$ is obtained by evaluating
\begin{equation}
\sigma = \frac{D - B}{\epsilon \times N_\mathrm{target} \times \Phi}
\end{equation}
where $D$ is the number of selected data events ($D=729$), $B$ is the predicted number of background after the constraint ($B=132$), $\epsilon$ is the efficiency ($\epsilon= 0.367$), $N_\mathrm{targets}$ is the number of scattering targets in the fiducial volume ($N_\text{targets} = 4.88 \times 10^{31}$ nucleons), and $\Phi$ is the energy-averaged $\nu_e+\bar\nu_e$ flux prediction over the fiducial volume given the total exposure ($\Phi = \qty{8.50e10}{\per\square\centi\meter}$).

The flux-averaged differential cross section is obtained by evaluating
\begin{equation}
\bigg\langle \frac{d\sigma}{dx}\bigg\rangle_i = \frac{\sum_j U_{ij} \cdot(D_j - B_j)}{N_\mathrm{target} \times \Phi \times \Delta x_i}
\end{equation}
where $D_j$ and $B_j$ are the number of selected data events and predicted background events after constraint in bin $j$, and $\Delta x_i$ is the width of bin $i$ for observed variable $x$. $U_{ij}$ is the unfolding matrix, which corrects for smearing effects and selection inefficiencies. 

This work uses the Wiener-SVD method~\cite{Tang:2017rob} to perform the cross section unfolding. This method combines the use of singular value decomposition (SVD) unfolding~\cite{Hocker:1995kb} and regularization with a Wiener filter to extract the estimate of the true signal. The total unfolded covariance matrices, the regularization matrices $A_C$ from the unfolding procedure, and the $\chi^2$ goodness of fit are provided in the Supplemental Material, Secs.~\ref{sec:sm-response}--\ref{sec:sm-chisq}~\cite{SBND:supplemental}. Validations of the unfolding method and CCBC have been performed using alternative simulations, including alternative neutrino interaction generators, to confirm that the results are robust to the choice of underlying model. 

The measured flux-averaged total cross section and differential cross sections are compared with predictions from the following generators: \texttt{GENIE v3.6.2} \texttt{Ar23\_20i\_00\_000} (\texttt{Ar23})~\cite{Andreopoulos:2009rq,GENIE:2021npt}, \texttt{GiBUU 2025} (\texttt{GiBUU})~\cite{Buss:2011mx,Soplin:2023nxg}, \texttt{NuWro 25.11.1} (\texttt{NuWro})~\cite{Golan:2012rfa,Golan:2012wx}, and \texttt{NEUT 6.1.4} (\texttt{NEUT})~\cite{Hayato:2009zz,Hayato:2021heg}. Each generator prediction is obtained using the analysis' signal definition. The predictions are processed through the \texttt{NUISANCE} framework~\cite{Stowell:2016jfr} for comparison. 
\paragraph{Results}
The flux-averaged total cross section is measured to be $[3.90 \pm 0.18\,\text{(stat)} \pm 0.41\,\text{(syst)}] \times 10^{-39}$ \unit{\square{\centi\meter}\per nucleon}, where the average predicted neutrino energy for the signal is $\langle E_\nu\rangle=1.8$\,GeV. A comparison of the measured total cross section with generator predictions is shown in Table~\ref{tab:xsec_comparison}. The unfolded differential cross section in electron energy and angle is presented in Fig.~\ref{fig:unfolded_xsec} alongside generator predictions. All differential generator predictions shown are smeared with the regularization matrix $A_C$.

\begin{table}[h]
\caption{Comparison of the measured flux-averaged total cross section with generator predictions.}
\label{tab:xsec_comparison}
\begin{ruledtabular}
\begin{tabular}{lccc}
\textrm{Generator} & \textrm{$\sigma$ ($\times 10^{-39}$ \unit{\square{\centi\meter}\per nucleon})} & \textrm{$\chi^2$/dof} & \textrm{$p$-value} \\
\colrule
\texttt{NuWro 25.11.1}     & 4.37 & 1.07/1 & 0.301 \\
\texttt{GENIE 3.6.2}       & 4.49 & 1.66/1 & 0.198 \\
\texttt{NEUT 6.1.4}        & 4.77 & 3.69/1 & 0.055 \\
\texttt{GiBUU 2025}        & 5.03 & 6.30/1 & 0.012 \\
\colrule
This measurement & \multicolumn{3}{c}{$3.90 \pm 0.18\,\text{(stat)} \pm 0.41\,\text{(syst)}$} \\
\end{tabular}
\end{ruledtabular}
\end{table}

Given the $\chi^2$/dof and $p$-values shown in Table~\ref{tab:xsec_comparison}, the data indicates a preference for the \texttt{GENIE\:Ar23} and \texttt{NuWro} predictions, which predict a smaller overall cross section compared to the \texttt{GiBUU} and \texttt{NEUT} generators. 

We see signs of potential mismodeling in the highest-energy bin and the forward-most-direction bin, but this is largely within the uncertainties for \texttt{NuWro}, \texttt{GENIE Ar23}, and \texttt{NEUT}. For \texttt{GiBUU}, the tension is more significant, but this is driven largely by the overall normalization rather than the shape, as reflected in Table~\ref{tab:xsec_comparison}. A free-floating normalization yields $\chi^2\mathrm{/dof}=8.2/4$ and $p\mathrm{-value}=0.085$ in electron direction for \texttt{GiBUU}, indicating consistency in the shape between data and generator predictions.

In the relevant lepton energy range ($E_e > 500$ MeV), two previous exclusive differential measurements from MicroBooNE~\cite{MicroBooNE:2022tdd, MicroBooNE:2025aiw} also observe a model overprediction, but the overall results are well within statistical and systematic uncertainties. 
Where comparisons to the following generators are provided, the MicroBooNE inclusive NuMI results~\cite{MicroBooNE:2021gfj, MicroBooNE:2021ppm} also tend to prefer these generator predictions over \texttt{GiBUU} or \texttt{NEUT} predictions. The results from this work mirror the same preference, with increased significance, for a smaller total cross section across energy and angular bins. Higher statistics and future multi-differential measurements from SBND will more precisely probe these features. 

\paragraph{Conclusions}
This work presents measurements of the inclusive charged-current electron neutrino cross sections on argon, and the first electron neutrino cross section measurement with SBND. For electron energies above \qty{500}{\mega\eV}, we extract the total flux-averaged cross section and the flux-averaged differential cross sections as a function of electron energy and angle with respect to the beam axis. 
The performance of the selection and a comprehensive evaluation of the uncertainties are also reported. The measurements are compared to several neutrino event generators common in the field; due to the preference for overall lower cross section, they are in best agreement with the predictions from \texttt{GENIE} \texttt{Ar23} and \texttt{NuWro}.

The results presented in this work are the first measurements of a differential inclusive $\nu_e+\bar\nu_e$ cross section with the Booster Neutrino Beam, and are also the highest statistics measurements of electron neutrinos on argon to date. The energy regime probed by this work, with substantial sensitivity to electron neutrino energies above 1\,GeV, has significant overlap with the neutrino energy range relevant to the DUNE oscillation program~\cite{DUNE:2020jqi}. As such, these results provide a direct, data-driven measurement of $\nu_e$-Ar interaction models, especially important for reducing $\nu_e/\nu_\mu$ modeling uncertainties that impact DUNE’s sensitivity to $CP$ violation. Future measurements in SBND will lower the electron energy threshold and be able to leverage over 10 times the statistics used in this work, enabling multi-differential cross section measurements. Furthermore, improvements to the reconstruction, flux modeling, and detector modeling will improve the selection performance and further reduce the associated uncertainties on measurements of electron neutrinos in SBND.

\paragraph{Acknowledgments}
This document was prepared by the SBND Collaboration using the resources of the Fermi National Accelerator Laboratory (Fermilab), a U.S. Department of Energy, Office of Science, Office of High Energy Physics HEP User Facility. Fermilab is managed by FermiForward Discovery Group, LLC, acting under Contract No. 89243024CSC000002. The SBND Collaboration acknowledges the generous support of the following organizations: the U.S. Department of Energy, Office of Science, Office of High Energy Physics; the U.S. National Science Foundation; Los Alamos National Laboratory for LDRD funding, the Science and Technology Facilities Council (STFC), part of United Kingdom Research and Innovation (UKRI), the UKRI Future Leaders Fellowship (grant number MR/V022407/1), and The Royal Society; the Swiss National Science Foundation; the Spanish Ministerio de Ciencia, Innovacíon y Universidades (MICIU/ AEI/ 10.13039/ 501100011033) under grants No PRE2019-090468, CNS2022-136022, RYC2022-036471-I, PID2023-147949NB-C51 \& C53 and Comunidad de Madrid (PEJ-2023-AI/COM-28399); the European Union’s Horizon 2020 research and innovation program under GA no 101004761 and the Marie Sklodowska-Curie grant agreements No 822185, 101081478, and 101003460; the São Paulo Research Foundation 1098 (FAPESP), the National Council of Scientific and Technological Development (CNPq) and Ministry of Science, Technology \& Innovations-MCTI of Brazil; the Minas Gerais research foundation (FAPEMIG), grants APQ-00544-23 and APQ-01249-24; the Anusandhan National Research Foundation (ANRF, India) under the Ramanujan Fellowship (Grant No. RJF/2025/000203). An award of computer time was provided by the ASCR Leadership Computing Challenge (ALCC) program. This research used resources of the Argonne Leadership Computing Facility, which is a U.S. Department of Energy Office of Science User Facility operated under contract DE-AC02-06CH11357.


\bibliography{references}

\clearpage
\onecolumngrid
\setcounter{equation}{0}
\setcounter{figure}{0}
\setcounter{table}{0}
\setcounter{section}{0}
\setcounter{secnumdepth}{5}
\renewcommand{\theequation}{S\arabic{equation}}
\renewcommand{\thefigure}{S\arabic{figure}}
\renewcommand{\thetable}{S\arabic{table}}
\renewcommand{\thesection}{S-\Roman{section}}

\begin{center}
  {\large\textbf{Supplemental Material}}
\end{center}

\section{Fiducial Volume and Flux}
\label{sec:sm-flux}
The measurement presented is flux-averaged over the fiducial volume of the detector. To describe the fiducial volume used for this analysis, we define the coordinate system of the detector as follows: $\hat x$ in the drift direction with $x=0$ set at the cathode, $\hat y$ in the vertical direction, with $y=0$ at the vertical center, and $\hat z$ in the beam direction with $z=0$ at the upstream face. The fiducial volume of the detector is defined as \qty{10}{\centi\meter} away from the outer detector edges, \qty{50}{\centi\meter} away from the downstream detector wall, and \qty{5}{\centi\meter} away from the cathode. To ensure electric field uniformity, the fiducial volume additionally excludes the region defined by 100\,cm from the top face and 250\,cm from the downstream face for the right-hand drift volume. The fiducial volume definition can be summarized:

\begin{gather}
 |x| \in [5, 190] \text{ cm}, \\
 z \in [10, 450] \text{ cm}, \\
 y \in \begin{cases}
      [-190, 190] \text{ cm} & \text{if } z \in [10,250] \text{ cm for all } x \\
      [-190, 190] \text{ cm} & \text{if } z \in [250,450] \text{ cm and } x > 0 \\
      [-190, 100] \text{ cm} & \text{if } z \in [250,450] \text{ cm and } x < 0.
    \end{cases}
\end{gather}

The resulting $\nu_e$ and $\bar\nu_e$ flux spectrum is included in the data release and reported in units of neutrinos per cm$^2$ per POT per GeV. The flux is binned in 125 MeV intervals, covering the range from 0 to 10 GeV. 

\newpage
\section{Shower Calibration}
\label{sec:sm-shower-cal}
\begin{figure}[H]
    \centering
    \includegraphics[width=0.6\linewidth]{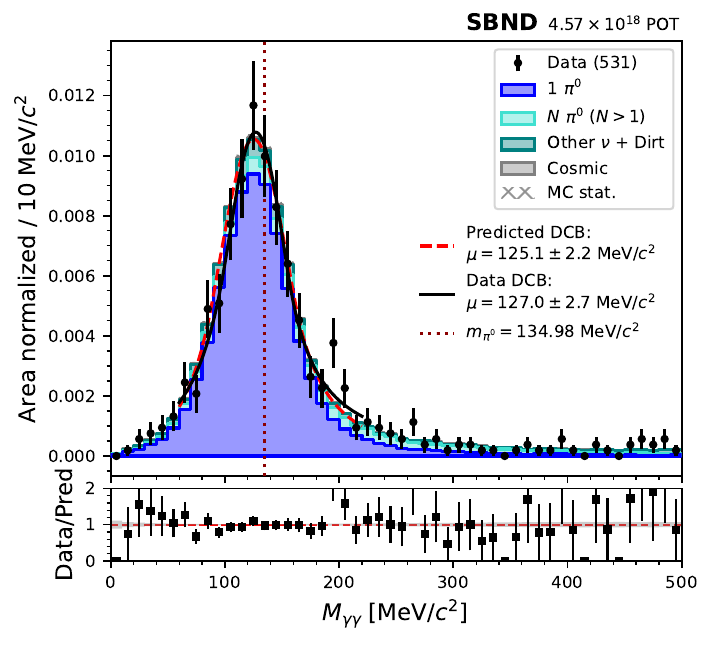}
    \caption{Reconstructed neutral pion $\pi^0$ mass peak distributions from selected photon-induced electromagnetic showers in data and simulation. A double-sided Crystal Ball function is shown for data and the prediction (black and red, respectively).}
    \label{fig:pi0_mass_peaks_gen1}
\end{figure}

Shower-energy correction factors are derived from simulation as a function of reconstructed energy by comparing the true and reconstructed energies of simulated photons. Overlapping 50 MeV-wide windows, spaced at 10 MeV intervals, are scanned across a 1 GeV reconstructed energy range. Within each window, the distribution of the ratio of true to reconstructed shower energy, \(E_{\mathrm{true}}/E_{\mathrm{reco}}\), is fit with a double-sided Crystal Ball function. The peak value extracted from each fit defines the correction factor for that window and is assigned to the mean reconstructed energy of the showers within the window. A smoothing spline is then fit to these values to obtain a continuous energy-dependent correction function. The resulting correction factor approaches a constant value of approximately 1.17 at high energies. As a validation of the correction, the reconstructed \(\pi^{0}\) mass distributions in simulation and a subset of the data are compared after applying the same correction procedure. As shown in Fig.~\ref{fig:pi0_mass_peaks_gen1}, the fitted mass peak is \(127.0 \pm 2.7\)\,MeV/c$^2$ in data and \(125.1 \pm 2.2\) MeV/c$^2$ in simulation. Due to the asymmetric, non-gaussian smearing in the energy reconstruction of individual EM showers, there remains a residual misalignment between the reconstructed and true $\pi^0$ mass peak.

\newpage
\section{Selection Performance}
\label{sec:sm-selection}
We provide further details on the selection steps and performance. Before applying selection cuts, a small number (1\%) of signal interactions are not reconstructed, leading to a 99\% starting efficiency. The neutrino candidate selection primarily rejects cosmic-ray and beam-induced background with a combination of scintillation light and TPC topology information. To reject muon neutrino interactions, we reject interactions where the leading track has a reconstructed length greater than 200\,cm. To select electron-induced electromagnetic showers and reject photon-induced electromagnetic showers, we apply cuts on the shower conversion gap (\textless 2\,cm) and the $dE/dx$ at the beginning of shower development (1.25$<dE/dx<$2.5\,{MeV/cm}). The final selection requirements reject the remaining background, typically mis-reconstructed particles, with selection cuts on shower development variables such as the opening angle ($\theta_\mathrm{open} < 0.5$\,rad) and length (\textless 200\,cm).
\begin{table}[H]
\caption{Overview of each step of the $\nu_e+\bar\nu_e$CC selection, with the efficiency and purity after each selection cut.}
\label{tab:selection}
\begin{ruledtabular}
\begin{tabular}{llcc}
Stage & Selection Cut & Efficiency (\%) & Purity (\%)\\
\colrule
Neutrino Candidate Selection & No Cuts                                & 99.0 & $<$0.5 \\
                             & Light Timing, PE, matching score       & 93.8 & $<$0.5 \\
                             & \texttt{Pandora} $\nu$-Cosmic BDT                       & 89.5 & $<$0.5 \\
                             & Vertex in FV                           & 88.4 & $<$0.5 \\
                             & High-Energy Shower                     & 79.1 & 13.2   \\
Muon Rejection               & Track Length                           & 78.0 & 14.7   \\
Electron Selection           & Shower conversion gap                  & 59.1 & 26.8   \\
                             & Shower dE/dx                           & 41.1 & 56.7   \\
                             & Shower opening angle                   & 36.8 & 78.5   \\
                             & Shower length                          & 36.7 & 80.1   \\
\end{tabular}
\end{ruledtabular}
\end{table}

\newpage
\section{$dE/dx$ of Reconstructed Showers}
\label{sec:sm-dedx}
\begin{figure}[H]
    \centering
    \includegraphics[width=0.6\linewidth]{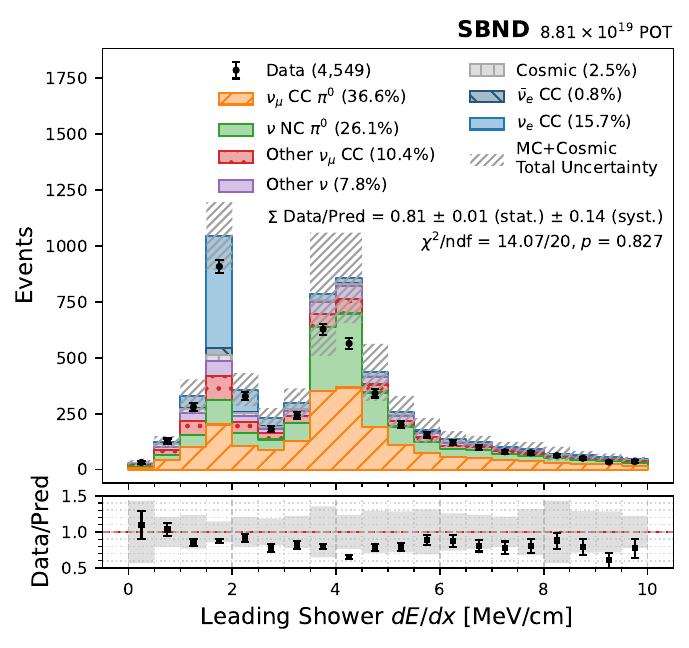}
    \caption{The $dE/dx$ of the leading reconstructed electromagnetic shower after applying the muon rejection cut.}
    \label{fig:dedx}
\end{figure}

The energy deposited per unit length, or $dE/dx$, of the first three centimeters of the reconstructed shower are shown in Fig.~\ref{fig:dedx}. The shower candidates shown include the neutrino candidate selection cuts, as well as the muon rejection cut.

\newpage
\section{Systematic Uncertainty Estimation}
\label{sec:sm-covariance}
\label{sec:sm-syst}
The covariance matrix for the systematic uncertainties is defined as
\begin{equation}
    \Cov(n_a, n_b) = \frac{1}{N_{\text{univ}}} \sum^{N_{\text{univ}}}_{u=1} \big(n_a^u - n_a^\cv\big) \big(n_b^u - n_b^\cv\big),
\end{equation}
where $n^u$ is the event rate in universe $u$, $n^\cv$ is the event rate in the central value simulation, and $a,b$ refer to reconstructed bins. The total covariance matrix is the sum of all systematic and statistical covariance matrices, where we assume Poisson uncertainties for the data statistical uncertainty.

\begin{figure}[H]
    \centering
    \includegraphics[width=0.75\linewidth]{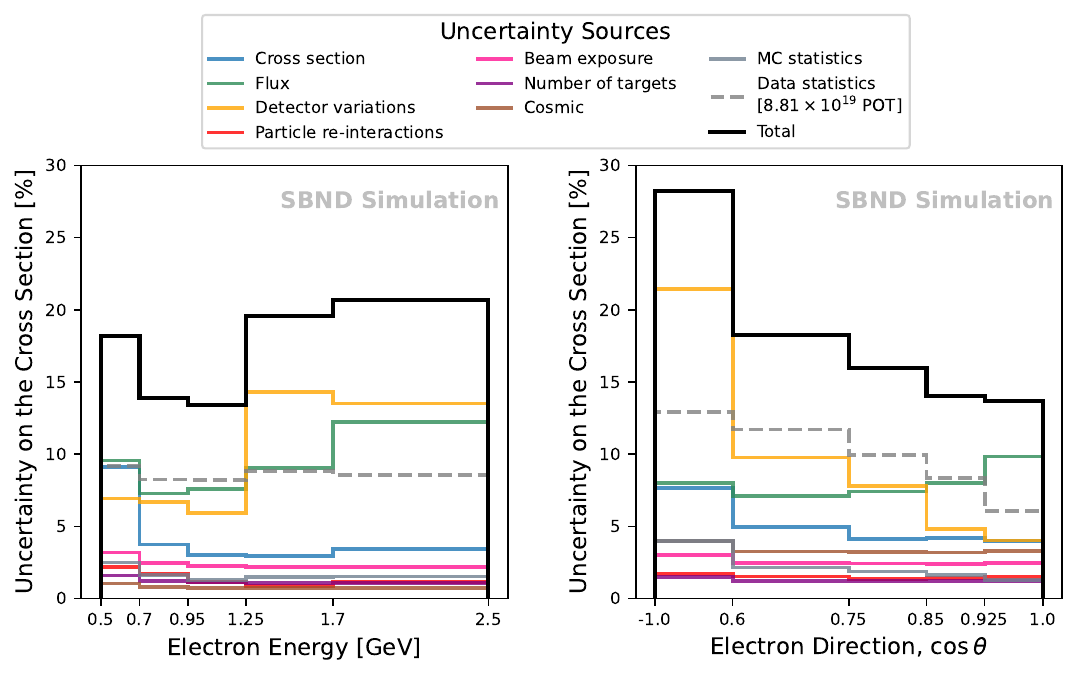}
    \caption{Fractional uncertainties on the cross section measurement as a function of electron energy (left) and direction $\cos\theta$ (right), separated by source.}
    \label{fig:syst_xsec}
\end{figure}

The fractional uncertainty per bin, separated by source, as a function of the electron energy and direction $\cos\theta$ is shown in Fig.~\ref{fig:syst_xsec}. Systematic sources of uncertainty include the cross section model, flux model, detector variations, particle re-interactions, exposure accounting, and cosmic-ray uncertainties. Statistical uncertainties on both the simulation and data are also included.

\newpage
\section{Sideband Selected Events}
\label{sec:sm-sideband}
\begin{figure}[H]
    \centering
    \includegraphics[width=\linewidth]{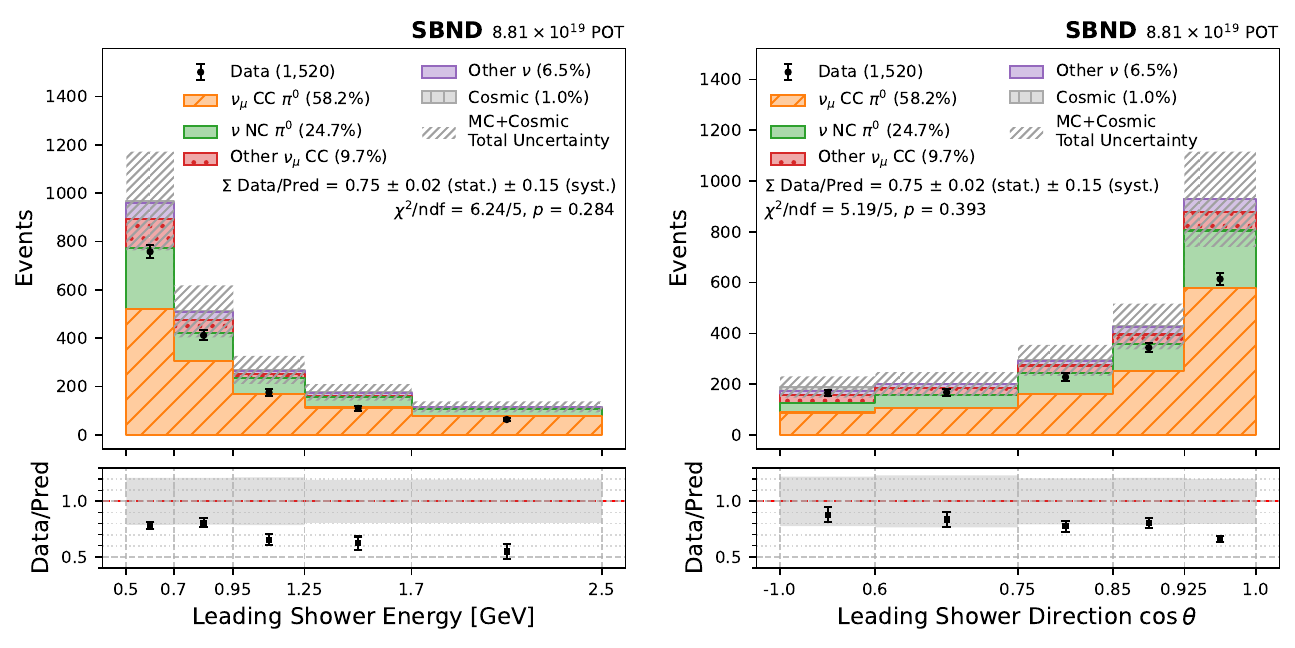}
    \caption{Event rate distributions of selected sideband events as a function of leading shower energy (left) and direction $\cos\theta$ (right).}
    \label{fig:sideband_evt_rate}
\end{figure}

The event rate distributions of selected sideband events are shown as a function of leading shower energy and direction $\cos\theta$ in Fig.~\ref{fig:sideband_evt_rate}.

\newpage
\section{Unconstrained Selected $\nu_e+\bar\nu_e$CC Event Rate}
\label{sec:sm-unconstr}
\begin{figure}[H]
    \centering
    \includegraphics[width=1\linewidth]{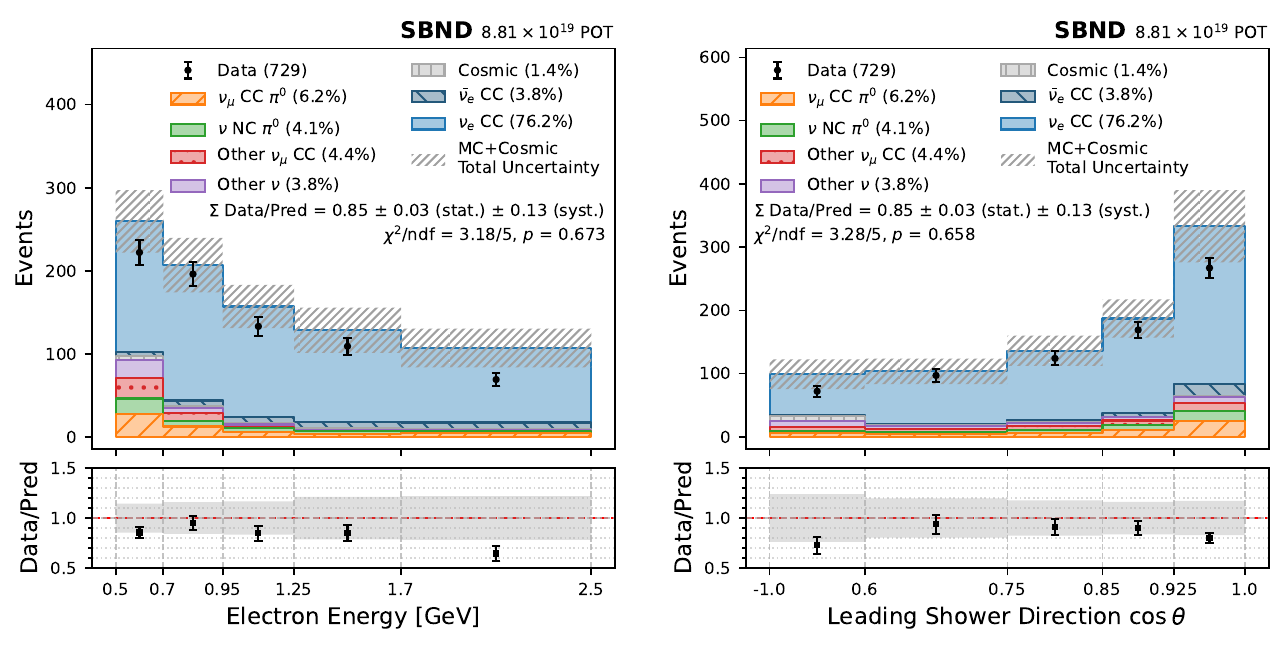}
    \caption{Selected signal event rate before applying background constraint in bins of electron energy (left) and direction $\cos\theta$ (right).}
    \label{fig:unconstr_evt_rate}
\end{figure}

The event rate of selected events before applying the background constraint is shown in Fig.~\ref{fig:unconstr_evt_rate}.

\newpage
\section{Detector Response Matrix}
\label{sec:sm-response}
\begin{figure}[H]
    \centering
    \includegraphics[width=\linewidth]{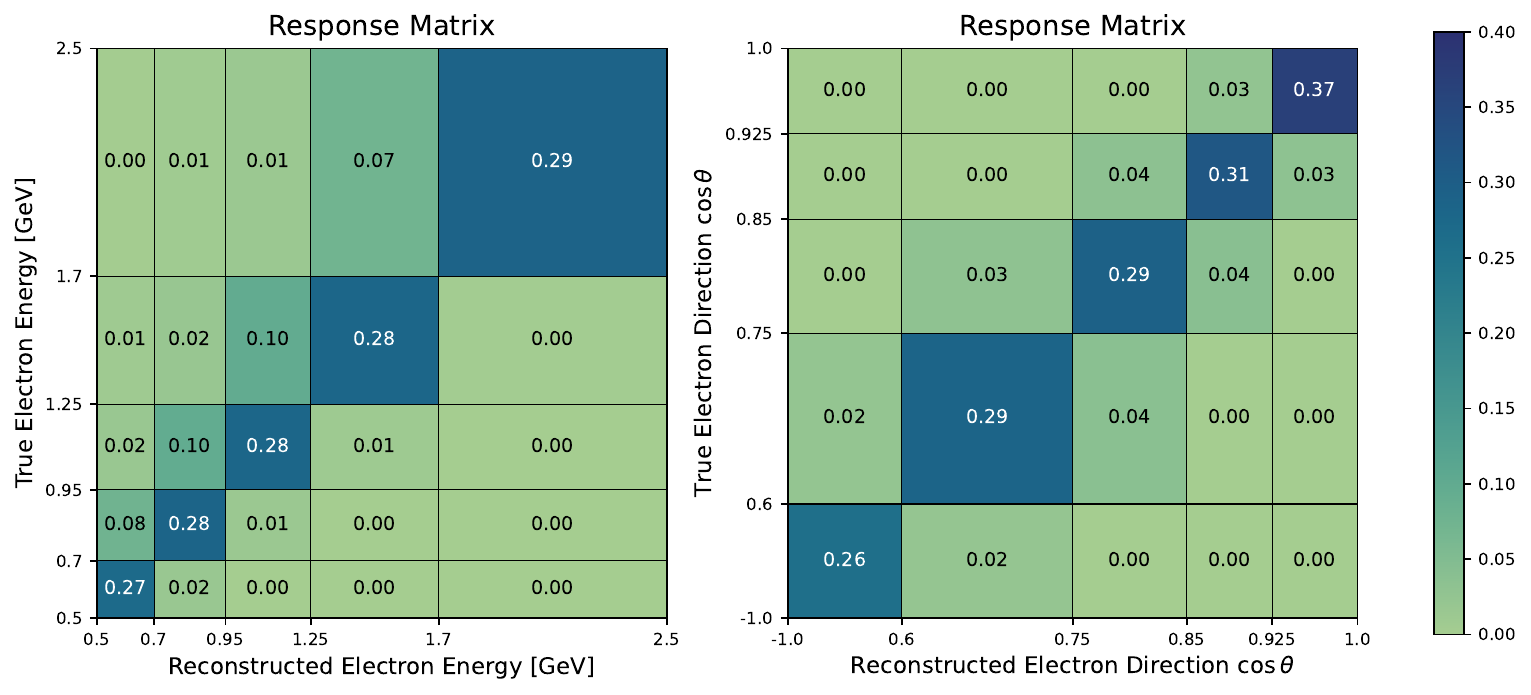}
    \caption{The detector response matrix as a function of electron energy (left) and direction $\cos\theta$ (right).}
    \label{fig:response_matrix}
\end{figure}

The detector response matrix is shown as a function of the electron energy and direction $\cos\theta$ in Fig.~\ref{fig:response_matrix}.

\section{Regularization Matrix}
\label{sec:sm-regularization}
\begin{figure}[H]
    \centering
    \includegraphics[width=\linewidth]{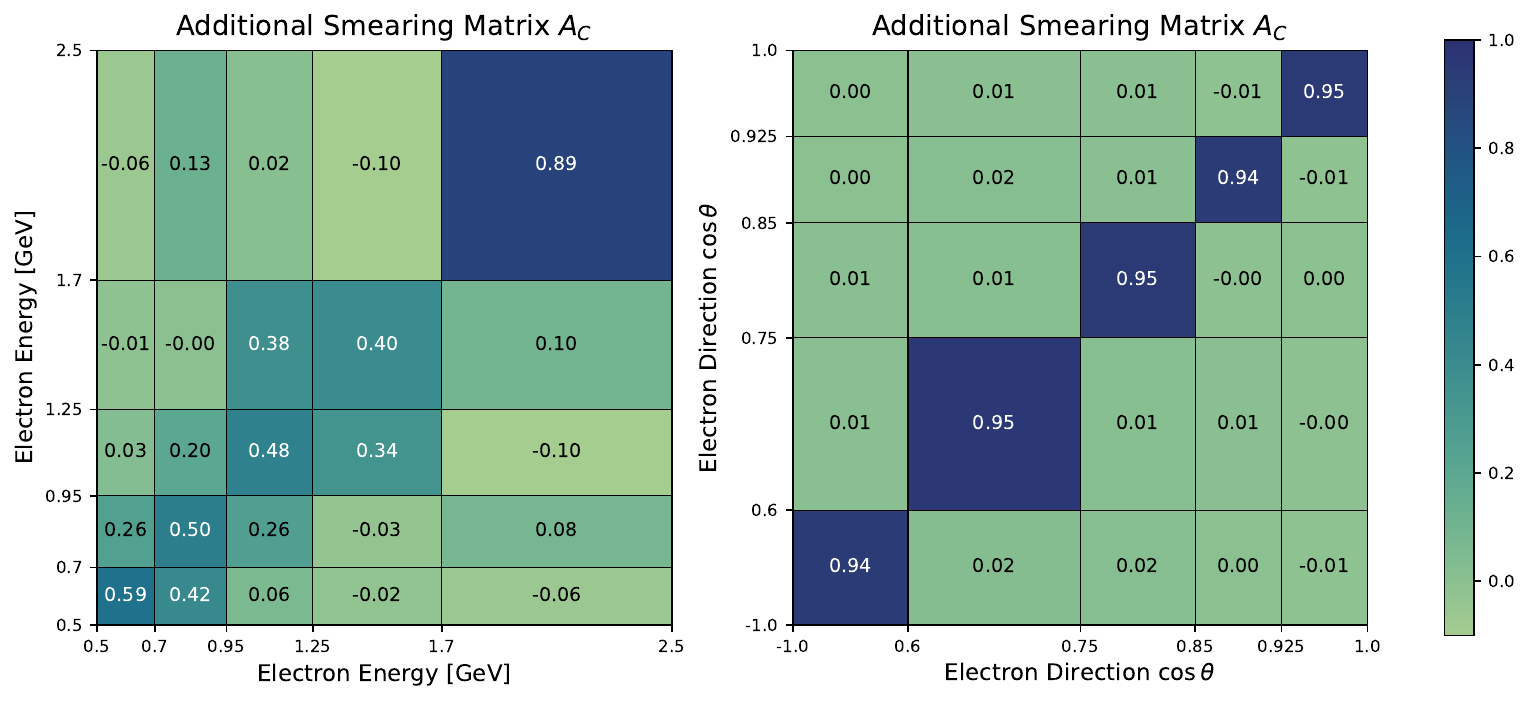}
    \caption{The regularization, or additional smearing matrix, $A_C$ as a function of electron energy (left) and direction $\cos\theta$ (right).}
    \label{fig:addsmear_matrix}
\end{figure}

The regularization matrix, or additional smearing matrix, $A_C$, from the Wiener-SVD unfolding method is shown as a function of the electron energy and direction $\cos\theta$ in Fig.~\ref{fig:addsmear_matrix}. These matrices should be applied to the true differential cross section predictions when comparing to the unfolded data differential cross section.

\section{Unfolded Covariance Matrix}
\label{sec:sm-unfolded-cov}
\begin{figure}[H]
    \centering
    \includegraphics[width=\linewidth]{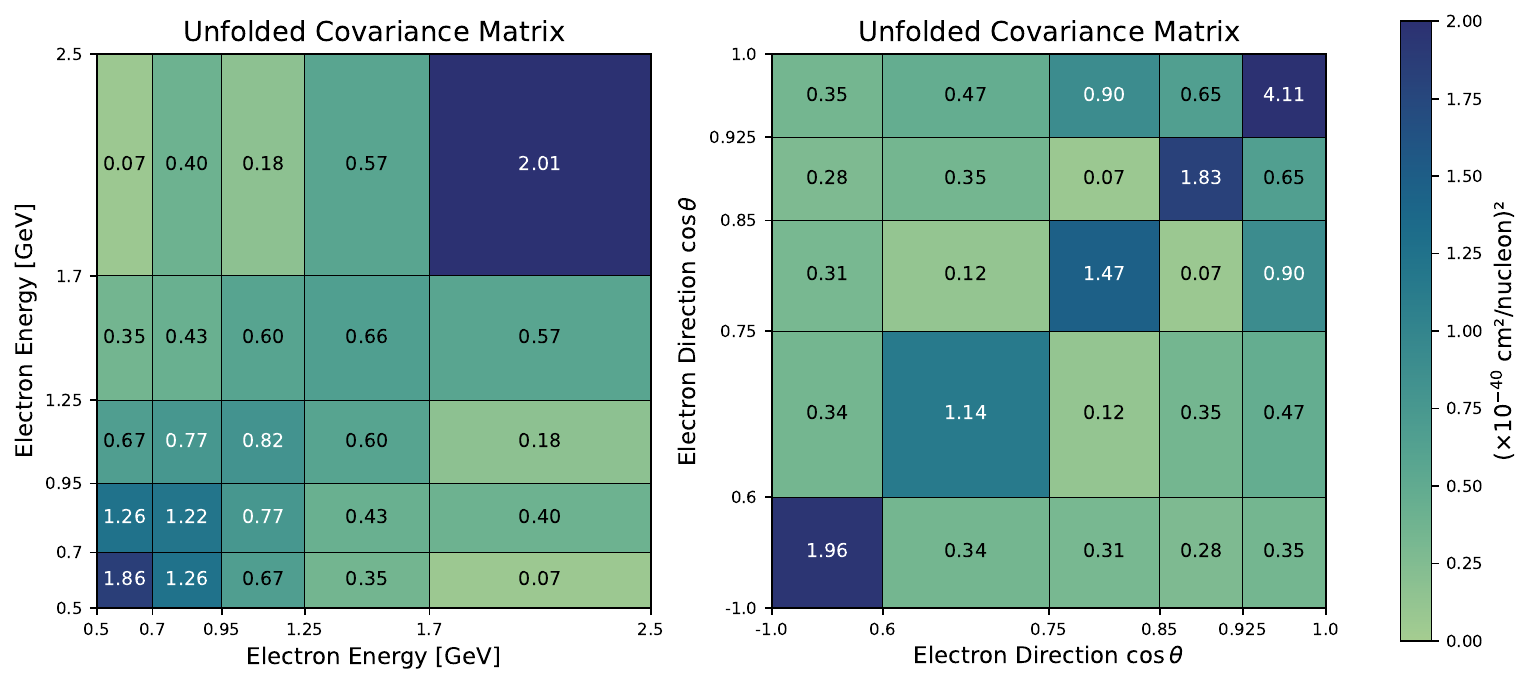}
    \caption{The unfolded covariance matrix as a function of the electron energy (left) and direction $\cos\theta$ (right).}
    \label{fig:unfolded_covariance}
\end{figure}

The total unfolded covariance matrix for the differential cross section as a function of the electron energy and direction $\cos\theta$ is shown in Fig.~\ref{fig:unfolded_covariance}. These matrices include statistical and systematic uncertainties.

\section{$\chi^2$ Goodness of Fit}
\label{sec:sm-chisq}
To quantify the goodness of fit between the unfolded data and generator predictions, we use a $\chi^2$ expressed as
\begin{equation}
    \label{eq:xsec-chisq_smeared}
    \chi^2 = (\hat\bphi- A_C \cdot \bphi)^T\cdot (\Cov^\mathrm{unf})^{-1} \cdot (\hat\bphi- A_C \cdot \bphi),
\end{equation}
where $\hat\bphi$ is the estimate of the true signal from data, $A_C$ is the regularization matrix, $\bphi$ is the signal prediction, and $\Cov^\mathrm{unf}$ is the unfolded covariance matrix.
\end{document}